\documentclass[aps,prb,reprint,twocolumn,superscriptaddress,floatfix,nofootinbib]{revtex4-1}
\usepackage{amsmath,amssymb,color,comment}
\usepackage{graphicx}
\usepackage[english]{babel}
\usepackage[bookmarks=true,colorlinks,linkcolor=blue,urlcolor=blue,citecolor=blue]{hyperref}
\DeclareMathOperator{\Tr}{Tr}
\usepackage{bm,ulem}
\usepackage{mathrsfs}
\usepackage{soul}
\usepackage{braket}

\newcommand{\rme}{{\rm e}}
\newcommand{\rmi}{{\rm i}}

\newcommand{\App}[1]{Appendix~\ref{#1}}

\def\d{\mathrm d}

\def\figpath{.}

\begin{document}

\title{Anomalous dynamical phase in quantum spin chains with long-range interactions}

\author{Ingo Homrighausen}
\affiliation{Universit\"at G\"ottingen, Institut f\"ur Theoretische Physik, Friedrich-Hund-Platz 1, 37077 G\"ottingen, Germany}

\author{Nils O.~Abeling}
\affiliation{Universit\"at G\"ottingen, Institut f\"ur Theoretische Physik, Friedrich-Hund-Platz 1, 37077 G\"ottingen, Germany}

\author{Valentin Zauner-Stauber}
\affiliation{Vienna Center for Quantum Technology, University of Vienna, Boltzmanngasse 5, 1090 Wien, Austria}

\author{Jad C.~Halimeh}
\affiliation{Physics Department and Arnold Sommerfeld Center for Theoretical Physics, Ludwig-Maximilians-Universit\"at M\"unchen, D-80333 M\"unchen, Germany}
\affiliation{National Institute for Theoretical Physics (NITheP), Stellenbosch 7600, South Africa}
\affiliation{Institute of Theoretical Physics, Department of Physics, University of Stellenbosch, Stellenbosch 7600, South Africa}

\date{\today}

\begin{abstract}
The existence or absence of non-analytic cusps in the Loschmidt-echo return rate is traditionally employed to distinguish between a regular dynamical phase (regular cusps) and a trivial phase (no cusps) in quantum spin chains after a global quench. However, numerical evidence in a recent study [J.~C.~Halimeh and V.~Zauner-Stauber, arXiv:1610.02019] suggests that instead of the trivial phase a distinct \textit{anomalous} dynamical phase characterized by a novel type of non-analytic cusps occurs in the one-dimensional transverse-field Ising model when interactions are sufficiently long-range. Using an analytic semiclassical approach and exact diagonalization, we show that this anomalous phase also arises in the fully-connected case of infinite-range interactions, and we discuss its defining signature. Our results show that the transition from the regular to the anomalous dynamical phase coincides with $\mathbb{Z}_2$-symmetry breaking in the infinite-time limit, thereby showing a connection between two different concepts of dynamical criticality. Our work further expands the dynamical phase diagram of long-range interacting quantum spin chains, and can be tested experimentally in ion-trap setups and ultracold atoms in optical cavities, where interactions are inherently long-range.
\end{abstract}
\maketitle

\section{Introduction}

Dynamical phase transitions have recently been the subject of intense theoretical and experimental investigation. Most commonly, they fall into two main types, both of which involve a quench where a control parameter in the system Hamiltonian is abruptly switched from some initial value to a final one, subsequently throwing the system out of equilibrium. The first kind of dynamical phase transition (DPT-I),\cite{Sciolla2011,Smacchia2015,Zunkovic2016a,Zunkovic2016b,Halimeh2016a} is of the Landau type: one waits for the system to relax into a (quasi-)steady state and extracts a suitable order parameter, usually that associated with spontaneous symmetry breaking in the system at equilibrium. This is done as a function of the final value of the quench-control parameter, and if a non-analyticity arises in this function, then a DPT-I has occurred in the system.

\begin{figure}[t]
 \centering
 \includegraphics[width=\linewidth]{\figpath/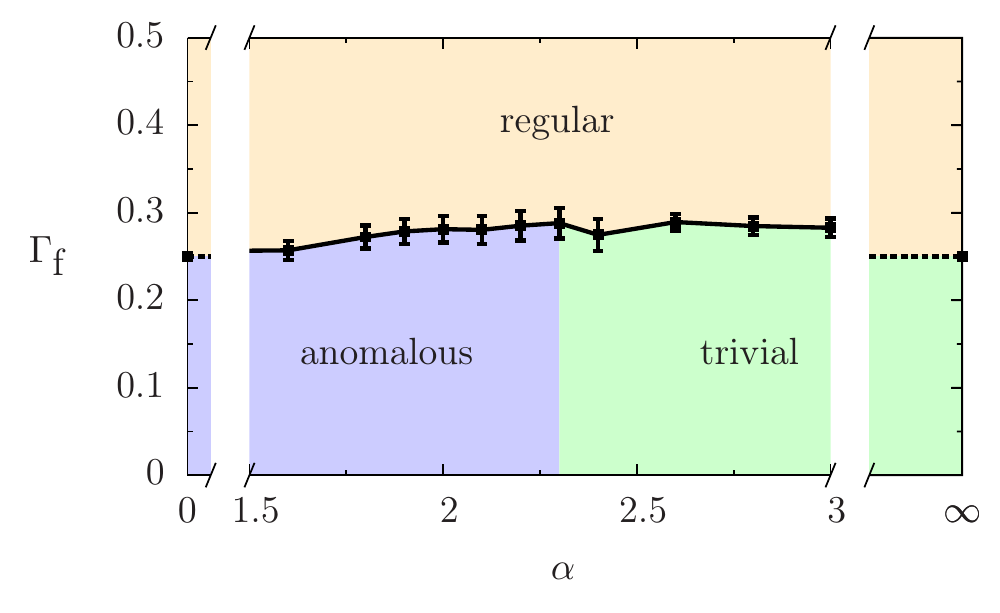}
 % Plot3_PhaseDiagram.pdf: 531x338 pixel, 72dpi, 18.73x11.92 cm, bb=
 \caption{(Color online) The dynamical phase diagram of the one-dimensional LR-TFIM~\eqref{eq:LRTFIM} after a global quench with initial field strength $\Gamma_{\text{i}}=0$, showing three distinct dynamical phases: regular, anomalous, and trivial (see main text). The dynamical critical line is marked in solid black. The results for the nonintegrable model are obtained using iMPS,\cite{Halimeh2016b} while the dynamical critical point for the NN-TFIM ($\alpha\to\infty$) is known analytically.\cite{Heyl2013} The phase diagram for the FC-TFIM ($\alpha=0$) is the main result of this work.
 }
 \label{fig:PhaseDiagram}
\end{figure}

A second type of dynamical phase transition is the DPT-II,\cite{Heyl2013,Heyl2014} in which non-analyticities in time, or lack thereof, in the Loschmidt-echo return rate

\begin{align}
r(t)=-\lim_{N\to\infty}\frac{1}{N}\ln|\braket{\psi_0|e^{-\rmi \hat{\mathcal{H}} t}|\psi_0}|^2,
\label{eq:ratefun}
\end{align}

\noindent characterize different phases, with pre-quench ground state $|\psi_0\rangle$, system size $N$, and post-quench Hamiltonian $\hat{\mathcal{H}}$. In the context of the DPT-II, an analogy\cite{Heyl2013} is made between the thermal partition function and the Loschmidt echo $\braket{\psi_0|e^{-\rmi \hat{\mathcal{H}} t}|\psi_0}$, or, equivalently, between the thermal free energy and the Loschmidt-echo return rate $r(t)$, where evolution time is now interpreted as a complex inverse temperature. Consequently, if the Loschmidt-echo return rate exhibits non-analyticities in evolution time after a quench, this is analogous to non-analyticities in the free energy of a system in equilibrium, which is the hallmark of an equilibrium phase transition.\cite{Footnote1} This DPT-II, first classified in the seminal work of Ref.~\onlinecite{Heyl2013} for the one-dimensional nearest-neighbor transverse-field Ising model (NN-TFIM), has been studied both analytically\cite{Pozsgay2013,Heyl2014,Hickey2014,Vajna2014,Vajna2015,Heyl2015,Schmitt2015,Campbell2016,Heyl2017,Zunkovic2016a,Piroli2017}and numerically\cite{Karrasch13,Fagotti2013,Canovi2014,Kriel2014,Andraschko2014,Sharma2015,Zhang2016,Sharma2016,Abeling2016,Vid2016,Zunkovic2016b,Halimeh2016b,Halimeh2017} in various models, and has also been experimentally observed.\cite{Peng2015,Flaeschner2016,Jurcevic2016} Even though for certain quenches \cite{Heyl2013} the critical final value of the quenching parameter that separates the phase with cusps from that with no cusps coincides with the equilibrium critical point of the model, this is not always the case,\cite{Andraschko2014,Vajna2014} and in general the \textit{dynamical critical point} separating such dynamical phases is different from its equilibrium counterpart.

In Fig.~\ref{fig:PhaseDiagram} we show, in the context of the DPT-II for quenches from zero field strength, the dynamical phase diagram of the one-dimensional long-range transverse-field Ising model (LR-TFIM) given by the Hamiltonian

\begin{align}\label{eq:LRTFIM}
\hat{\mathcal{H}}(\Gamma)=-\frac{J}{2\mathcal{N}}\sum_{i\neq j}^N\frac{1}{|i-j|^{\alpha}}\hat{S}^z_i\hat{S}^z_j-\Gamma\sum_i\hat{S}^x_i,
\end{align}

\noindent where $\hat{S}^a_i$, $a=x,y,z$, are the spin-$1/2$ operators on site $i$, $J>0$ is the spin-spin coupling constant, $\Gamma$ is the strength of the transverse magnetic field, $\alpha\geq 0$, and $\mathcal{N}$ is the Kac normalization\cite{Kac1963} given by

\begin{align}
\mathcal{N}=\frac{1}{N-1}\sum_{i\neq j}^N\frac{1}{|i-j|^{\alpha}}=\frac{2}{N-1}\sum_{n=1}^N\frac{N-n}{n^{\alpha}},
\end{align}

\noindent which guarantees energy-density intensivity for $\alpha\leq 1$. The part of this diagram at $\alpha=0$ is the main result of this work. The part of this phase diagram for $\alpha>1$ has been constructed using Matrix Product State (MPS) techniques for infinite systems, a method known as iMPS.\cite{Fannes1992,Verstraete2008,Haegeman2011,Haegeman2016,Stauber2017,Halimeh2016b}
In the limit $\alpha\to\infty$ the nearest-neighbor result\cite{Heyl2013} is obtained.
As can be seen in Fig.~\ref{fig:PhaseDiagram}, quenching from zero field strength to above a certain \textit{dynamical} critical value sets the system in a regular dynamical phase characterized by the appearance of an infinite sequence of cusps with the first cusp appearing before the first minimum in the Loschmidt-echo return rate.
These cusps become sharper and temporally less separated with increasing quench strength.\cite{Heyl2013,Halimeh2016b} However, for sufficiently long-range interactions ($\alpha\lesssim 2.3$), a new \textit{anomalous} dynamical phase\cite{Halimeh2016b} appears whose defining signature is that cusps appear only after the first minimum in the return rate. In contrast to their regular counterparts, the anomalous cusps separate less in time from each other with decreasing quench strength, with more smooth maxima emerging in the return rate before their onset. In fact, numerical results\cite{Halimeh2016b} suggest that these cusps arise for arbitrarily small quenches, even though in the framework of iMPS and time-dependent density matrix renormalization group\cite{White1992,Schollwoeck2005,Schollwoeck2011,White2004,Verstraete2004,Vidal2004,Daley2004,Gobert2005} ($t$-DMRG) techniques entanglement buildup prevents access to long-enough evolution times that would be necessary to see the onset of these anomalous cusps for extremely weak quenches.

In this paper, we turn our attention to the analytically-tractable fully-connected transverse-field Ising model (FC-TFIM), and investigate the nature of the anomalous phase in a semiclassical approach.\cite{Sciolla2011} The advantage of this is two-fold: (i) In iMPS, it is intrinsically difficult to include the Kac-normalization to ensure intensivity of the energy density for $\alpha\leq 1$, whereas the FC-TFIM allows for investigating the anomalous phase with exact diagonalization (ED) and semiclassical techniques. ED is a technique which is fundamentally different from iMPS methods, therefore it additionally provides an alternate venue to study the anomalous phase. (ii) Moreover, from an intuitive point of view, it is logical to consider the limit of infinite-range interactions since it appears that the anomalous phase occurs only for interactions that are sufficiently long-range.

The remainder of the paper is organized as follows: In Sec.~\ref{sec:FC-TFIM} we review the FC-TFIM and use a semiclassical treatment to derive the infinite-time average of the $\mathbb{Z}_2$ order parameter and its oscillation period. In Sec.~\ref{sec:results} we present and discuss our results obtained from ED, characterize the anomalous phase, and discuss the connection between the cusps in the return rate and the $\mathbb{Z}_2$ order parameter. We conclude in Sec.~\ref{sec:Conc}.

\section{Fully-connected transverse-field Ising model}\label{sec:FC-TFIM}
\subsection{Model and quench}
The one-dimensional FC-TFIM is described by taking the $\alpha=0$ limit of~\eqref{eq:LRTFIM},

\begin{align}
\label{eq:hamiltonian_fully_connected}
\hat{\mathcal{H}}(\Gamma)=-\frac{J}{2N}\sum_{i\neq j}^N\hat{S}^z_i\hat{S}^z_j-\Gamma\sum_i\hat{S}^x_i-\epsilon\sum_i\hat{S}^z_i.
\end{align}

\noindent where we have additionally introduced a $\mathbb{Z}_2$-symmetry-breaking term with $\epsilon$ a small positive longitudinal field of $\mathscr{O}(1/N)$, because we 
treat finite-size systems only and spontaneous symmetry breaking is a feature of the thermodynamic limit. The FC-TFIM has an equilibrium quantum critical point\cite{Das2006} at $\Gamma_{\text{c}}^{\text{e}}=J/2$.
Hence, in the ground state of (\ref{eq:hamiltonian_fully_connected}) the longitudinal magnetization is positive for $\Gamma<\Gamma_{\text{c}}^{\text{e}}$ and vanishes for $\Gamma>\Gamma_{\text{c}}^{\text{e}}$.

We are interested in the DPT-II and its corresponding dynamical phases in the FC-TFIM whilst using $\Gamma$ as the quench-control parameter. In the following, we shall prepare our system in the ground state $|\psi_0\rangle$ of $\hat{\mathcal{H}}(\Gamma_{\text{i}})$, and then at time $t=0$, the field strength is suddenly switched from $\Gamma_{\text{i}}$ to $\Gamma_{\text{f}}\neq\Gamma_{\text{i}}$, leading to time-evolving the system under $\hat{\mathcal{H}}(\Gamma_{\text{f}})$ and subsequently discerning from the return rate what dynamical phase our system is in from the perspective of the DPT-II. The DPT-I in this model was first studied in Ref.~\onlinecite{Sciolla2011}. Moreover, it was argued that there is an equivalence\cite{Zunkovic2016b,Halimeh2016b} between the DPT-I and DPT-II in the LR-TFIM, and also in the FC-TFIM.\cite{Zunkovic2016b} Nevertheless, to the best of our knowledge, the anomalous phase has not been previously investigated outside of Ref.~\onlinecite{Halimeh2016b}, which does so numerically in the context of the LR-TFIM for $\alpha>1$.

\subsection{Semiclassical equations of motion}\label{sec:semiclassics}
\begin{figure}[t]
 \centering
 \includegraphics[width=\linewidth]{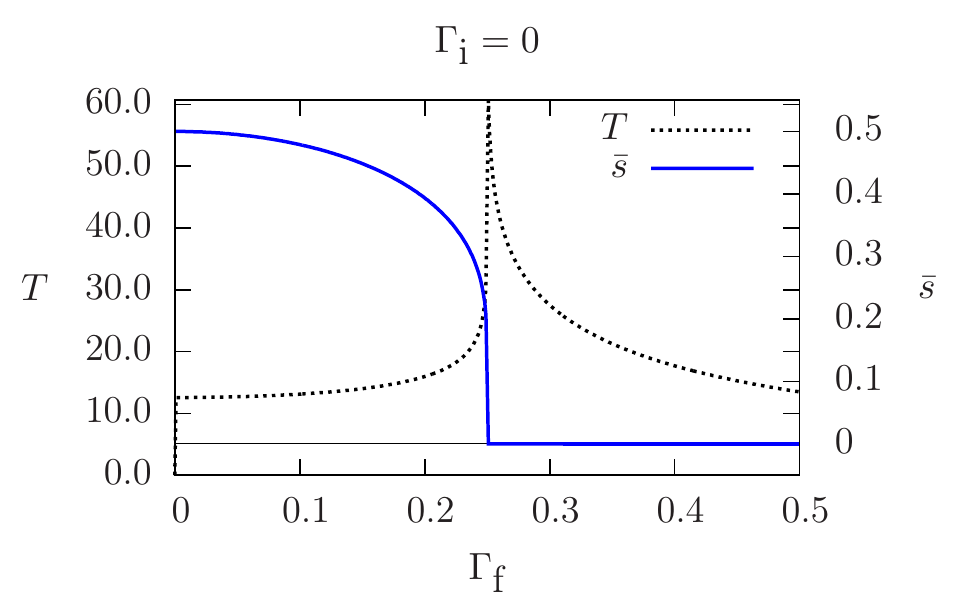}
 % Plot3_PhaseDiagram.pdf: 531x338 pixel, 72dpi, 18.73x11.92 cm, bb=
 \caption{(Color online). The periodicity of the order parameter (dotted black line) in the FC-TFIM for a quench from $\Gamma_{\text{i}}=0$ to $\Gamma_{\text{f}}$, derived in a semiclassical approach. This periodicity is also that of the non-analytic cusps arising in the Loschmidt-echo return rate~\eqref{eq:ratefun}.
The period diverges at the dynamical critical point $\Gamma^{\text{d}}_{\text{c}}=0.25$, i.e.~at the point where the infinite time average of the longitudinal magnetization (solid blue line) is non-analytic as a function of $\Gamma_{\text{f}}$.
The critical point $\Gamma^{\text{d}}_{\text{c}}$ also separates the anomalous and regular phases.}
%  In agreement with what has been previously shown in the LR-TFIM using $t$-DMRG\cite{Halimeh2016a} and iMPS,\cite{Halimeh2016b} $\Gamma^{\text{d}}_{\text{c}}<\Gamma_{\text{c}}^{\text{e}}$.}
 \label{fig:period}
\end{figure}

The period of the $\mathbb{Z}_2$ order parameter $\hat{s}^z(t)=\sum_i\hat{S}^z(t)/N$ can be computed in an effective semiclassical picture.\cite{Sciolla2011} To leading order in the mean-field limit $N\to \infty$, the post-quench magnetization expectation value $\langle\hat{s}^z\rangle=s(t)+\mathscr{O}(1/N)$ evolves according to Hamilton's equations of motion, $\dot{s}(t)=\partial_pH_{\text{eff}}$ and $\dot{p}(t)=-\partial_sH_{\text{eff}}$, with the effective Hamiltonian
\begin{align}\label{eq:eff_Hamiltonian}
H_{\text{eff}}(s,p)=-\frac{J}{2}s^2-\frac{\Gamma_{\text{f}}}{2}\sqrt{1-4s^2}\cos p,
\end{align}
and initial condition
\begin{align}\label{eq:InitCond}
s(0)&=
\begin{cases}
    0,& \text{if } \Gamma_{\text{i}}>\Gamma_{\text{c}}^{\text{e}},\\   \sqrt{\frac{1}{4}-\Gamma_{\text{i}}^2},& \text{if } \Gamma_{\text{i}}<\Gamma_{\text{c}}^{\text{e}},
\end{cases}\\[1em]
p(0)&=
\begin{cases}
    0,& \text{if } \Gamma_{\text{i}}\neq0,\\   -\pi/2,& \text{if } \Gamma_{\text{i}}=0.
\end{cases}
\end{align}
Henceforth, we choose units of time in which $J = 1$. The period of the classical orbit is
\begin{align}\nonumber
T &= 2\int_{s_-}^{s_+}\frac{\d s}{\partial_p H_{\text{eff}}}\\[1em]\label{eq:period}
&=2\int_{s_-}^{s_+}\frac{\d s}{\sqrt{\left(\frac{1}{4}-s^2\right)\Gamma_{\text{f}}^2-\left(E+\frac{1}{2}s^2\right)}},
\end{align}
and the average magnetization along this orbit is
\begin{align}\nonumber
\bar{s}&= \frac{1}{T}\int_0^Ts(t)\d t\\[1em]\label{eq:average}
&=\frac{2}{T}\int_{s_-}^{s_+}\frac{s\,\d s}{\sqrt{\left(\frac{1}{4}-s^2\right)\Gamma_{\text{f}}^2-\left(E+\frac{1}{2}s^2\right)}},
\end{align}
where the integration bounds $s_-<s_+$ are the turning points of the trajectory $s(t)$, and the energy,
\begin{align}
E=H_{\text{eff}}(s(0),p(0)),
\end{align}

\noindent is conserved. For $\Gamma<\Gamma_{\text{c}}^{\text{e}}$ the Hamiltonian~(\ref{eq:eff_Hamiltonian}) has a hyperbolic fixed point at $(s,p)=(0,0)$, whose stable directions are connected to the unstable directions by two homoclinic orbits.
%The period of these homoclinic orbits diverges.
The homoclinic orbits separate closed $\mathbb{Z}_2$-invariant orbits (i.e.~orbits that are invariant under $s\mapsto-s$) from closed orbits that are not $\mathbb{Z}_2$-invariant.
As pointed out in Ref.~\onlinecite{Sciolla2011}, this leads to a DPT-I at 

\begin{align}\label{eq:Gamma_d}
\Gamma^{\text{d}}_{\text{c}}(\Gamma_{\text{i}})=\left(\Gamma^{\text{e}}_{\text{c}}+\Gamma_{\text{i}}\right)/2.
\end{align}

\noindent For quenches to $\Gamma_{\text{f}}=\Gamma^{\text{d}}_{\text{c}}$ the initial condition (\ref{eq:InitCond}) lies on a homoclinic orbit and $s(t)$ approaches $s=0$ exponentially in time, i.e.~the period (\ref{eq:period}) of $s(t)$ diverges at $\Gamma^{\text{d}}_{\text{c}}$ as shown in Fig~\ref{fig:period}). For quenches to $\Gamma_{\text{f}}>\Gamma^{\text{d}}_{\text{c}}$ the orbit is $\mathbb{Z}_2$-symmetric and $s(t)$ oscillates around zero such that the infinite-time average 

\begin{align}
\bar{s}=\lim_{t\to\infty}\lim_{N\to\infty}\frac{1}{Nt}\int_0^t\d t'\;\sum_i\langle\hat{S}^z_i(t')\rangle
\end{align}

\noindent vanishes. Note that the limit $N\to\infty$ has to be taken before the limit $t\to\infty$ in order to obtain the semiclassical result~\eqref{eq:average}. In contrast, for $\Gamma_{\text{f}}<\Gamma^{\text{d}}_{\text{c}}$, the orbit is not $\mathbb{Z}_2$-symmetric and the infinite-time average takes a nonzero value, cf.~Fig.~\ref{fig:period}.

\section{Results and discussion}\label{sec:results}
\begin{figure*}[t]
 \centering
 \includegraphics[width=\linewidth]{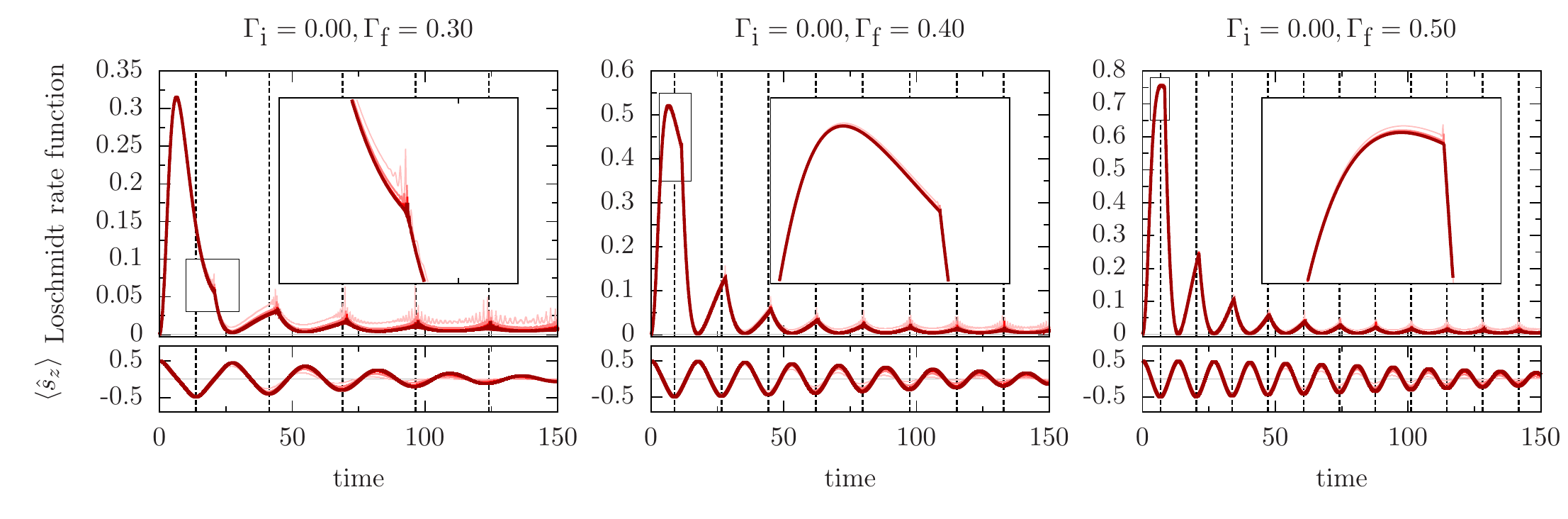}
%  \includegraphics{anomalous.pdf}
 % Plot3_PhaseDiagram.pdf: 531x338 pixel, 72dpi, 18.73x11.92 cm, bb=
 \caption{(Color online). Loschmidt-echo rate function and expectation value of the magnetization after a quench from $\Gamma_{\text{i}}=0$ to $\Gamma_{\text{f}}=0.30$ (left), $\Gamma_{\text{f}}=0.40$ (middle), and $\Gamma_{\text{f}}=0.50$ (right).
 All quenches are in the regular phase ($\Gamma_{\text{f}}>\Gamma^{\text{d}}_{\text{c}}=0.25$), compare Fig.~\ref{fig:anomalous} for quenches in the anomalous phase.
 Each plot shows four different system sizes, $N=200,400,600,800$, from light to dark red, with the latter achieving convergence for the results shown here.
The dotted grid indicates the turning points of $\langle s_z\rangle$ in the thermodynamic limit according to~\eqref{eq:period}.}
 \label{fig:regular}
\end{figure*}

\begin{figure*}[t]
 \centering
 \includegraphics[width=\linewidth]{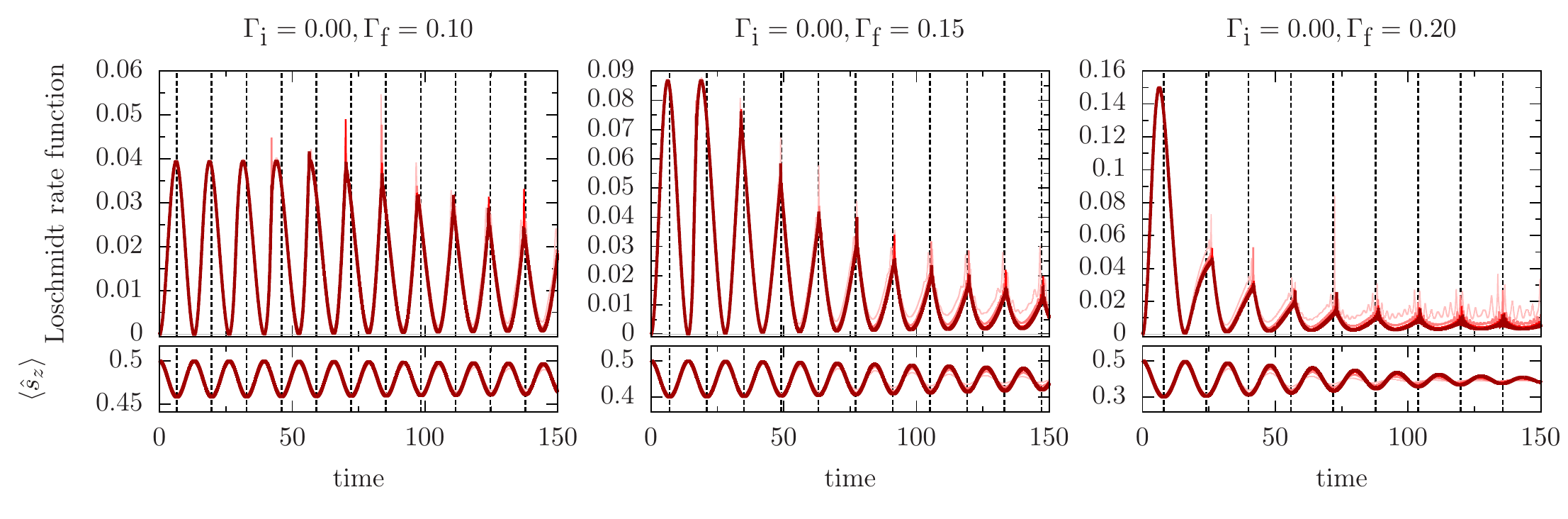}
%  \includegraphics{anomalous.pdf}
 % Plot3_PhaseDiagram.pdf: 531x338 pixel, 72dpi, 18.73x11.92 cm, bb=
 \caption{(Color online). Loschmidt-echo rate function and expectation value of the magnetization after a quench from $\Gamma_{\text{i}}=0$ to $\Gamma_{\text{f}}=0.10$ (left), $\Gamma_{\text{f}}=0.15$ (middle), and $\Gamma_{\text{f}}=0.20$ (right).
 All quenches are in the anomalous phase ($\Gamma_{\text{f}}<\Gamma^{\text{d}}_{\text{c}}=0.25$), compare Fig.~\ref{fig:regular} for quenches in the regular phase.
 Each plot shows four different system sizes, $N=200,400,600,800$, from light to dark red, with the latter achieving convergence for the results shown here.
The dotted grid indicates the turning points of $\langle s_z\rangle$ in the thermodynamic limit according to~\eqref{eq:period}.}
 \label{fig:anomalous}
\end{figure*}
We shall now present our results on the two distinct phases (regular and anomalous) of the DPT-II in the FC-TFIM, and argue that they are intimately related to the phases of the DPT-I in this model through sharing the same critical point $\Gamma^{\text{d}}_{\text{c}}$. Traditionally, the DPT-II is known to give rise to two phases: one with (regular) cusps for quenches across the DPT-II critical point, and a second with no cusps in the return rate for quenches not crossing it. In Ref.~\onlinecite{Heyl2013}, this was demonstrated in the case of the NN-TFIM, where it can be analytically shown that the DPT-II critical point is $\Gamma^{\text{e}}_{\text{c}}$. Much like the case of the NN-TFIM, the return rate in the FC-TFIM also shows regular cusps for quenches across $\Gamma^{\text{d}}_{\text{c}}$, as shown in Fig.~\ref{fig:regular} for $\Gamma_{\text{i}}=0$. In agreement with previous results,\cite{Heyl2013,Zunkovic2016b,Halimeh2016b} these cusps occur before the first minimum of the return rate and beyond. Also, the period of these cusps matches that of the order parameter at longer times and decreases with quench strength while the cusps themselves get sharper.

In the case of the NN-TFIM, cusps in the return rate are absent\cite{Heyl2013} for quenches below $\Gamma^{\text{e}}_{\text{c}}$, and the return rate is fully analytic. This has also been observed in Ref. \onlinecite{Halimeh2016b} to be the case for the LR-TFIM with sufficiently short-range interactions $\alpha\gtrsim2.3$. However, for longer-range interactions, the return rate does exhibit a new kind of cusps that are qualitatively different in their behavior from their regular counterparts. These cusps characterize the \textit{anomalous} dynamical phase, defined by a Loschmidt-echo return rate that displays non-analyticities only after its first minimum. In fact, it can be shown in iMPS that these anomalous cusps are caused by level crossings within the set of dominant eigenvalues of the MPS transfer matrix, which is qualitatively different from the set responsible for the manifestation of the regular cusps and which is dominant for quenches above the DPT-II critical point. In good agreement with iMPS data for the LR-TFIM, our ED results in Fig.~\ref{fig:anomalous} for the FC-TFIM show such anomalous cusps in the return rate for quenches below $\Gamma^{\text{d}}_{\text{c}}$, which, unlike the case of the NN-TFIM, is not equal to $\Gamma^{\text{e}}_{\text{c}}$ for the FC-TFIM. At longer times they also possess the same period as the order parameter, and, in contrast to the regular cusps, their period increases with quench strength. Moreover, they separate less in time and are preceded by more smooth maxima in the return rate with decreasing quench strength.

However, it is to be emphasized that the distinctive signature of the anomalous phase is that its cusps are delayed in the sense that they always occur after the first minimum of the return rate. This leads to smooth peaks preceding them, with more such analytic peaks the smaller the quench is. This can be seen in Fig.~\ref{fig:anomalous}, and agrees with what is observed in iMPS for the nonintegrable model\cite{Halimeh2016b} for $\alpha\lesssim2.3$. Additionally, we find that the anomalous cusps occur for arbitrarily small quenches in the FC-TFIM. 

\begin{figure}[t]
 \centering
 \includegraphics[width=0.75\linewidth]{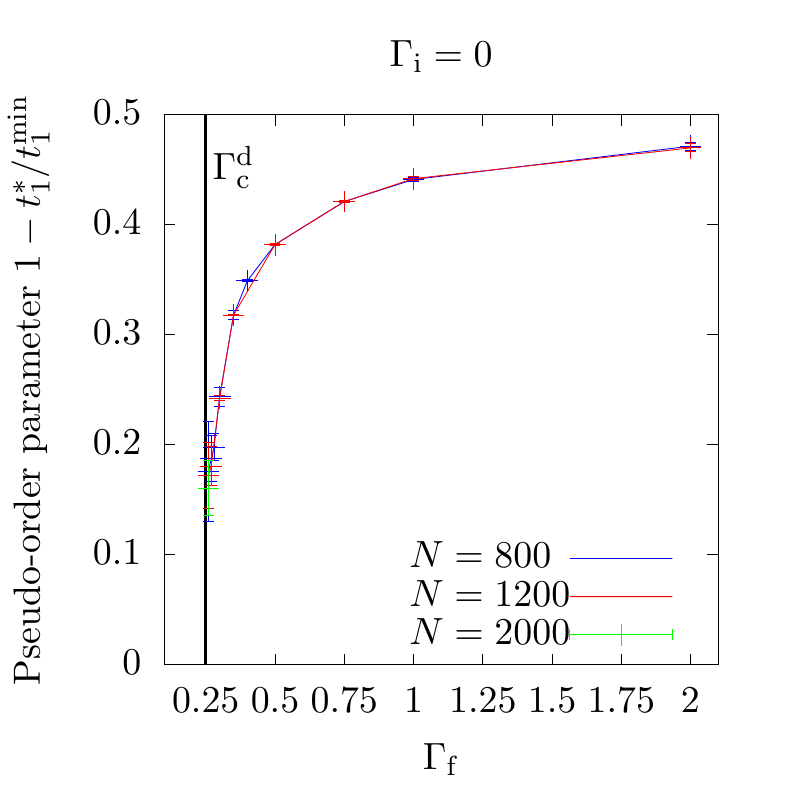}
%  \includegraphics{anomalous.pdf}
 % Plot3_PhaseDiagram.pdf: 531x338 pixel, 72dpi, 18.73x11.92 cm, bb=
 \caption{(Color online). The (pseudo-)order parameter $\eta$~\eqref{eq:pop} that tracks the transition from the regular phase ($\Gamma_{\text{f}}>\Gamma^{\text{d}}_{\text{c}}$) to the anomalous phase ($\Gamma_{\text{f}}<\Gamma^{\text{d}}_{\text{c}}$) for quenches from $\Gamma_{\text{i}}=0$ with system sizes $N=800,1200,2000$. $\eta$ decreases rapidly as the final value of the quench parameter approaches $\Gamma^{\text{d}}_{\text{c}}$. Even though in the thermodynamic limit this $\eta$ would exhibit sharp decay to zero, this is not expected to be the case in the finite systems we are able to simulate.}
 \label{fig:pop}
\end{figure}

The transition from the regular phase to the anomalous phase can be understood by observing the regular cusp before the first minimum of the return rate in each panel of Fig.~\ref{fig:regular}. This cusp moves away from the first maximum of the return rate and closer to the first minimum as $\Gamma_{\text{f}}$ is decreased towards $\Gamma^{\text{d}}_{\text{c}}$. Once $\Gamma_{\text{f}}\leq\Gamma^{\text{d}}_{\text{c}}$, this cusp crosses the first minimum of the return rate as we enter the anomalous phase, cf.~Fig.~\ref{fig:anomalous}. In fact, one can assign for quenches to $\Gamma_{\text{f}}>\Gamma_{\text{c}}^{\text{d}}$ a (pseudo-)order parameter\cite{Footnote2}

\begin{align}\label{eq:pop}
\eta=1-t_1^*/t_1^{\text{min}},
\end{align}

\noindent with $t_1^*$ the time at which the first cusp occurs and $t_1^{\text{min}}$ the time of the first minimum in the return rate. Fig.~\ref{fig:pop} shows this parameter decaying towards zero as one approaches the dynamical critical point from deep in the regular phase. For $\Gamma_{\text{f}}>>\Gamma_{\text{c}}^{\text{d}}$, we find that $\eta\to0.5$, meaning that the first cusp becomes situated exactly at the first maximum of the return rate, which is typical of quenches deep into the regular phase. As we are dealing with a finite system, this parameter will nevertheless not decay sharply to zero at $\Gamma_{\text{c}}^{\text{d}}$. In the thermodynamic limit, on the other hand, $\eta$ is expected to sharply decay to zero at $\Gamma_{\text{c}}^{\text{d}}$, but this limit is not accessible in our numerical simulations. As per definition, $\eta=0$ for $\Gamma_{\text{f}}<\Gamma_{\text{c}}^{\text{d}}$ because there is no cusp before the first minimum of the return rate in the anomalous phase. More details on the regular-to-anomalous transition are provided in~\App{sec:transition}.

It is evident in Figs.~\ref{fig:regular} and~\ref{fig:anomalous} that in the regular phase $\mathbb{Z}_2$ symmetry is preserved whereas in the anomalous phase it is broken with a non-vanishing average of the order parameter, in agreement with the infinite-time limit of Fig.~\ref{fig:period}. This indicates that the DPT-I and DPT-II are intimately related by sharing a common critical point $\Gamma^{\text{d}}_{\text{c}}$. Also, Figs.~\ref{fig:regular} and~\ref{fig:anomalous} indicate that the period of the cusps in either dynamical phase and that of the oscillations of the order parameter are the same at long times. In fact, our simulations show that the period of the cusps also grows indefinitely as $\Gamma_{\text{f}}\approx\Gamma^{\text{d}}_{\text{c}}$, in accordance with the diverging period of the order parameter shown in Fig.~\ref{fig:period}. As exemplified in \App{sec:NonzeroInitialField}, all findings also hold for other initial conditions $\Gamma_{\text{i}}\neq0$.

Furthermore, we comment that unlike in the LR-TFIM for $\alpha\lesssim2.3$ in Ref.~\onlinecite{Halimeh2016b}, the Loschmidt-echo return rate in the case of the FC-TFIM does not exhibit double-cusp structures. We speculate that these double cusps may be related to the nonintegrability of the LR-TFIM, and would thus be missing in the case of the FC-TFIM. We leave this question open for future investigation.

Finally, we remark that our ED results were extensively tested for convergence on various environments and using different independent implementations. In cases where the Loschmidt echo is very small, i.e. for large system sizes and at times when the Loschmidt return rate is large, we observed that double-precision ($\approx16$ significant digits) ED is not sufficient to numerically resolve the Loschmidt return rate. In order to get rid of the numerical noise, we performed the numerical computations with enhanced precision of up to $256$ significant digits.

\section{Conclusion}\label{sec:Conc}
Using semiclassical equations of motion and exact diagonalization, we have shown that the fully-connected transverse-field Ising model exhibits two distinct dynamical phases, one of which seems to occur as a direct result of the long-range interactions in this model. Starting in a $\mathbb{Z}_2$-symmetry-broken ground state, quenches below the dynamical critical point give rise to the \textit{anomalous} phase, whose defining signature is the occurrence of cusps only after the first minimum of the Loschmidt-echo return rate. On the other hand, quenches above the dynamical critical point lead to the regular phase, which shows cusps also before the first minimum of the return rate. The periods of the cusps in both phases display an intimate connection to the period of the $\mathbb{Z}_2$ order parameter oscillations. In fact, our ED simulations indicate that the anomalous phase coincides with the DPT-I phase of broken $\mathbb{Z}_2$ symmetry, while the regular phase with the DPT-I disordered phase. Our results agree with numerical results on the nonintegrable transverse-field Ising model with long-range interactions, obtained using an infinite matrix product state technique. Additionally, they provide support for the notion that long-range interactions bring about a new anomalous dynamical phase not found in short-range quantum spin chains. Our findings further extend the dynamical phase diagram of quantum spin chains with $\mathbb{Z}_2$ symmetry, and are suitable for investigation in ion-trap and optical cavity atom-photon experiments where interactions are long-range.

\section*{Acknowledgments}
The authors acknowledge discussions with Markus Heyl, Stefan Kehrein, and Ulrich Schollw\"ock, and are grateful to Michael Kastner for carefully reading and providing valuable comments on the manuscript.
Financial support from the Deutsche Forschungsgemeinschaft (DFG) through SFB/CRC1073 (Projects B03 and C03) is gratefully acknowledged.
V.~Z.-S. gratefully acknowledges support from the Austrian Science Fund (FWF): F4104 SFB ViCoM and F4014 SFB FoQuS.

\appendix

\begin{figure}[t]
 \centering
 \includegraphics[width=\linewidth]{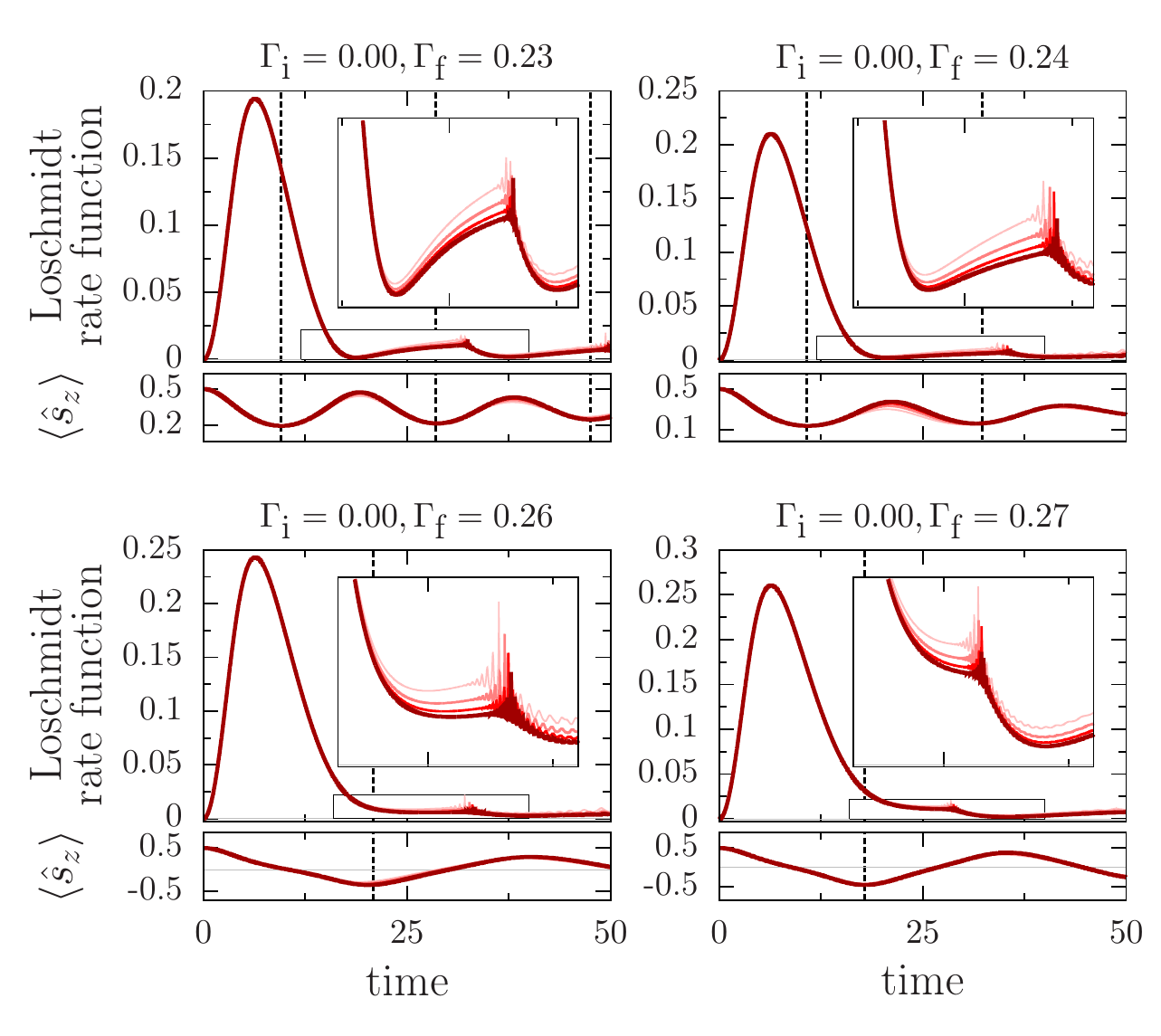}
%  \includegraphics{anomalous.pdf}
 % Plot3_PhaseDiagram.pdf: 531x338 pixel, 72dpi, 18.73x11.92 cm, bb=
 \caption{(Color online). Loschmidt-echo rate function and expectation value of the magnetization after a quench from $\Gamma_{\text{i}}=0$ to $\Gamma_{\text{f}}=0.23$ (top left), $\Gamma_{\text{f}}=0.24$ (top right), $\Gamma_{\text{f}}=0.26$ (bottom left), and $\Gamma_{\text{f}}=0.27$ (bottom right). Quenches in the top (bottom) panels are in the anomalous (regular) phase, $\Gamma_{\text{f}}<\Gamma^{\text{d}}_{\text{c}}=0.25$ ($\Gamma_{\text{f}}>\Gamma^{\text{d}}_{\text{c}}=0.25$). Each plot shows four different system sizes, $N=600,800,1000,1200$, from light to dark red.
The dotted grid indicates the turning points of $\langle s_z\rangle$ in the thermodynamic limit according to~\eqref{eq:period}.}
 \label{fig:transition}
\end{figure}

\section{Transition from anomalous to regular phase}

\label{sec:transition}
As mentioned in the main text, the transition from the anomalous to the regular phase manifests in the presence of a cusp immediately preceding the first minimum of the return rate in time at $\Gamma_{\text{f}}\gtrsim\Gamma^{\text{d}}_{\text{c}}$. This cusp then moves away from the first minimum to smaller times towards the first maximum of the return rate as one quenches deeper into the regular phase. Fig.~\ref{fig:transition} shows this behavior in the vicinity of $\Gamma^{\text{d}}_{\text{c}}$. For quenches very close to, yet below $\Gamma^{\text{d}}_{\text{c}}$ (top panels of Fig.~\ref{fig:transition}), the first cusp always appears after the first minimum of the return rate, which is the defining signature of the anomalous phase. However, for quenches right above $\Gamma^{\text{d}}_{\text{c}}$ (bottom panels of Fig.~\ref{fig:transition}), we see that the first cusp is no longer preceded by a minimum in the return rate, which defines the regular phase. Also to be noted is that, in agreement with the main results of Figs.~\ref{fig:regular} and~\ref{fig:anomalous}, Fig.~\ref{fig:transition} shows that the anomalous phase is linked to a finite nonzero average of the $\mathbb{Z}_2$ order parameter, while in the regular phase this order parameter vanishes.

Ideally, one would want to scan even closer to $\Gamma^{\text{d}}_{\text{c}}$, but this requires impracticable computational resources. The reason is that close to $\Gamma^{\text{d}}_{\text{c}}$ finite-size effects are particularly pronounced and one has to use large $N$ in order to see converged results.
This can be understood from the semiclassical picture discussed in Sec.~\ref{sec:semiclassics}.
For quenches close to $\Gamma^{\text{d}}_{\text{c}}$ the initial wave packet is localized near the homoclinic orbit of (\ref{eq:eff_Hamiltonian}) (recall that for the quench to $\Gamma_{\text{f}}=\Gamma^{\text{d}}_{\text{c}}$ the wave packet is exactly centered on the homoclinic orbit).
As time evolves the wave packet remains localized and follows the homoclinic orbit until it reaches the neighborhood of the unstable hyperbolic fixed point at $(s,p)=(0,0)$.
Even though the wave packet is not centered exactly at the hyperbolic point the wave packet's finite width of $\mathscr{O}(1/\sqrt{N})$ makes it `feel' the unstable directions.
As a consequence, the wave packet gets deformed and spreads in the unstable directions.
This leads to a deviation from the $N\to\infty$ result where the width of the wave packet remains localized also close to the hyperbolic point.
The closer one quenches to $\Gamma^{\text{d}}_{\text{c}}$, i.e.~the closer the wave packet comes to the hyperbolic fixed point, the larger $N$ has to be to avoid these finite-size effects. Thus, even though for the main results of the paper $N=800$ leads to convergence, for the quenches in this Appendix we have to go to larger $N$ to suppress most finite-size effects.

\section{Quenches from $\Gamma_{\text{i}}=0.20$}

\label{sec:NonzeroInitialField}
\begin{figure}[t]
 \centering
 \includegraphics[width=\linewidth]{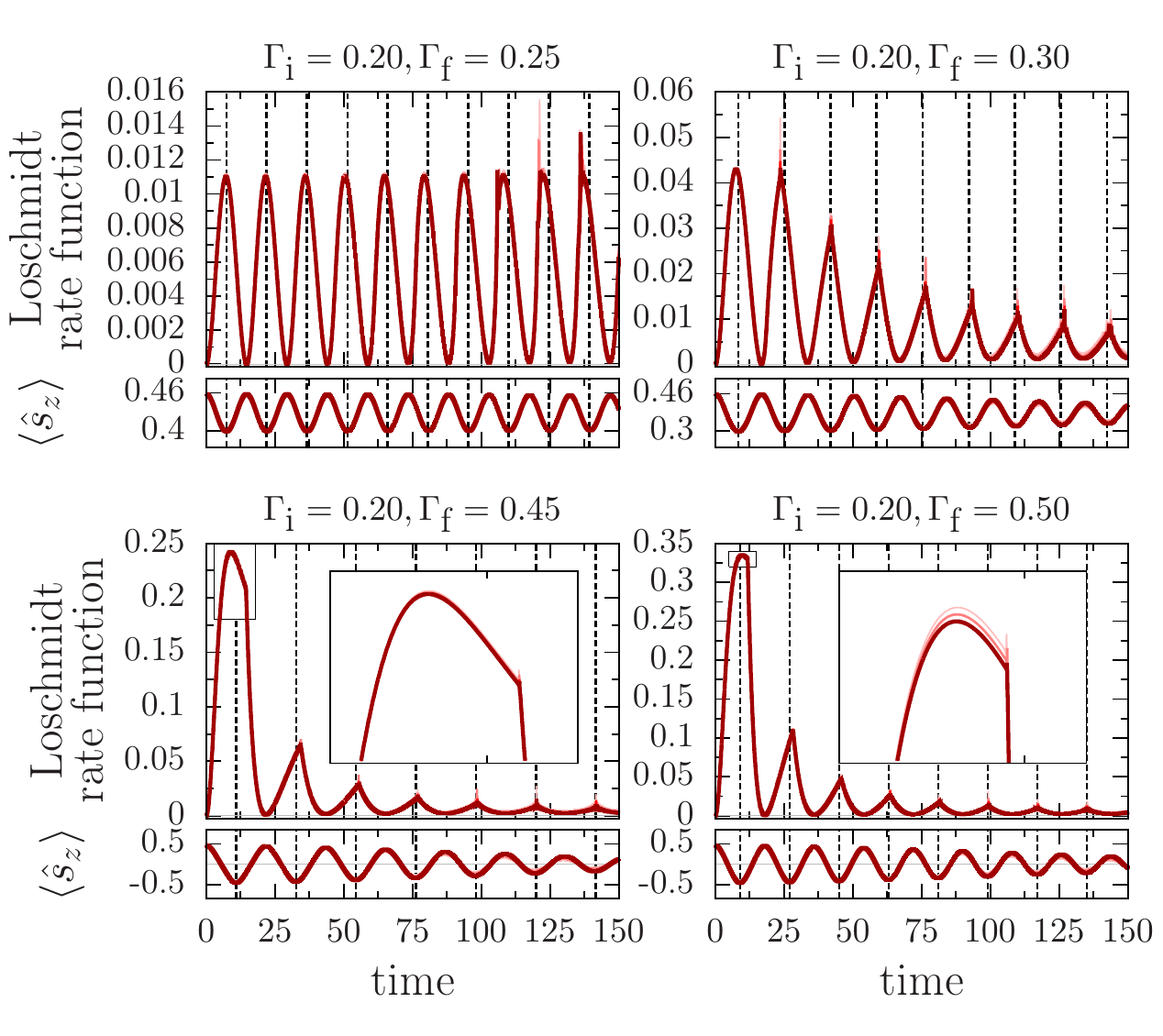}
%  \includegraphics{anomalous.pdf}
 % Plot3_PhaseDiagram.pdf: 531x338 pixel, 72dpi, 18.73x11.92 cm, bb=
 \caption{(Color online). Loschmidt-echo rate function and expectation value of the magnetization after a quench from $\Gamma_{\text{i}}=0.20$ to $\Gamma_{\text{f}}=0.25$ (top left), $\Gamma_{\text{f}}=0.30$ (top right), $\Gamma_{\text{f}}=0.45$ (bottom left), and $\Gamma_{\text{f}}=0.50$ (bottom right). Quenches in the top (bottom) panels are in the anomalous (regular) phase, $\Gamma_{\text{f}}<\Gamma^{\text{d}}_{\text{c}}=0.35$ ($\Gamma_{\text{f}}>\Gamma^{\text{d}}_{\text{c}}=0.35$). Each plot shows four different system sizes, $N=600,800,1000,1200$, from light to dark red, with the latter achieving convergence for the results shown here.
The dotted grid indicates the turning points of $\langle s_z\rangle$ in the thermodynamic limit according to~\eqref{eq:period}.}
 \label{fig:appendixB}
\end{figure}

We now look at the effect of changing the initial condition of our quench. Whereas the main part of the paper treats the case $\Gamma_{\text{i}}=0$, quenches with different initial values of the transverse-field strength lead to the same phase diagram with the only difference being quantitative because $\Gamma^{\text{d}}_{\text{c}}$ is a function of $\Gamma_{\text{i}}$ as expressed in~\eqref{eq:Gamma_d}. Nevertheless, the anomalous (regular) phase still manifests for quenches below (above) $\Gamma^{\text{d}}_{\text{c}}$. As an example, Fig.~\ref{fig:appendixB} shows four quenches from initial field strength $\Gamma_{\text{i}}=0.20$. For this initial value of the transverse field, the dynamical critical point according to~\eqref{eq:Gamma_d} is $\Gamma^{\text{d}}_{\text{c}}=0.35$ rather than $0.25$ when $\Gamma_{\text{i}}=0$ (see main results). In Fig.~\ref{fig:appendixB} we go from the anomalous phase (top panels) to the regular phase (bottom panels), where we see that in the anomalous phase the first cusp always occurs after the first minimum of the return rate, which is the defining feature of this phase. Note that the weaker the quench is in this phase, the more smooth maxima (and therefore the more smooth minima) precede the first cusp in time. However, after the transition to the regular phase, we see that the first cusp occurs before the first minimum of the return rate, which is the defining feature of this phase. This is qualitatively the same behavior as in the case of $\Gamma_{\text{i}}=0$ in the main part of the paper.

Additionally, Fig.~\ref{fig:appendixB} shows that the anomalous (regular) phase coincides with the $\mathbb{Z}_2$-symmetry-broken (unbroken) phase of the DPT-I for the case of $\Gamma_{\text{i}}=0.20$. This is also in agreement with our results in Figs.~\ref{fig:regular},~\ref{fig:anomalous}, and~\ref{fig:transition} for quenches from $\Gamma_{\text{i}}=0$.

%%%%%%%%%% Merge with supplemental materials %%%%%%%%%%

%--------------------------------------------------------

\begin{thebibliography}{9}
\bibitem{Footnote2}
One can in principle argue that $\eta$ is a sort of \cite{disorder parameter}, since it arises only in the regular phase of the DPT-II, which, at least for the FC-TFIM coincides with the disordered phase of the DPT-I.
\end{thebibliography}
\end{document}